\documentclass[nature,pdflatex,sn-mathphys-num]{sn-jnl}

\usepackage{graphicx,epsfig,psfrag,amsfonts,latexsym,eucal,amsmath,amsthm,color,amsbsy,float,amssymb,lineno,comment,dsfont,multicol,bm,mathrsfs,mathtools,calc,lmodern,subcaption,fix-cm}
\def\dd{\mbox{d}}

\begin{document}

\title[Article Title]{Revealing  wave-wave resonant interactions in ocean wind waves}

\author[1]{\fnm{Davide} \sur{Maestrini}}\email{damaestr@gmail.com}

\author[1,2]{\fnm{Giovanni} \sur{Dematteis}}\email{giovannidematteis@gmail.com}

\author[3]{\fnm{Alvise} \sur{Benetazzo}}\email{	alvise.benetazzo@cnr.it}

\author[1,4]{\fnm{Miguel} \sur{Onorato}}\email{miguel.onorato@unito.it}

\affil[1]{\orgdiv{Dipartimento di Fisica}, \orgname{Università di Torino}, \orgaddress{\street{Via P. Giuria 1}, \city{Torino}, \postcode{10125}, \country{Italy}}}

\affil[2]{\orgdiv{Physical Oceanography Department}, \orgname{Woods Hole Oceanographic Institution}, \orgaddress{\street{360 Woods Hole Rd}, \city{Woods Hole, MA}, \postcode{02543}, \country{USA}}}

\affil[3]{\orgdiv{Institute of Marine Sciences}, \orgname{CNR-ISMAR}, \orgaddress{\city{Venice}, \country{Italy}}}

\affil[4]{\orgdiv{INFN}, \orgname{Sezione di Torino}, \orgaddress{\street{Via P. Giuria 1}, \city{Torino}, \postcode{10125}, \country{Italy}}}



\abstract{Ocean wind waves are a fundamental manifestation of complex dynamics in geophysical fluid systems, characterized by a rich interplay between dispersion and nonlinearity. While linear wave theory provides a first-order description of wave motion, real-world oceanic environments are governed by nonlinear interactions that are responsible for a transfer of energy between waves of different lengths. Established theoretical concepts predict that four-wave resonant interactions serve as the primary mechanism for energy transfers among wave components in oceanic surface wave fields. Although the presence and efficiency of these resonant interactions have been demonstrated in controlled wave tank experiments, their direct identification in the real ocean, where a large number of random waves interact, has remained elusive. Here, using a stereoscopic system that enables the measurement of surface elevation in both space and time, we provide experimental evidence of resonant interactions in ocean wind waves. Our data not only reproduce the well-known figure-eight pattern predicted by Phillips, but also reveal a continuum of different resonant configurations that closely match the theoretical predictions. These findings support the validity of third-generation ocean wave models, strengthening their ability to accurately capture wave dynamics in the ocean.
}


\maketitle

\section*{Introduction}\label{sec1}
\begin{figure}
\centering
\includegraphics[width=1\linewidth]{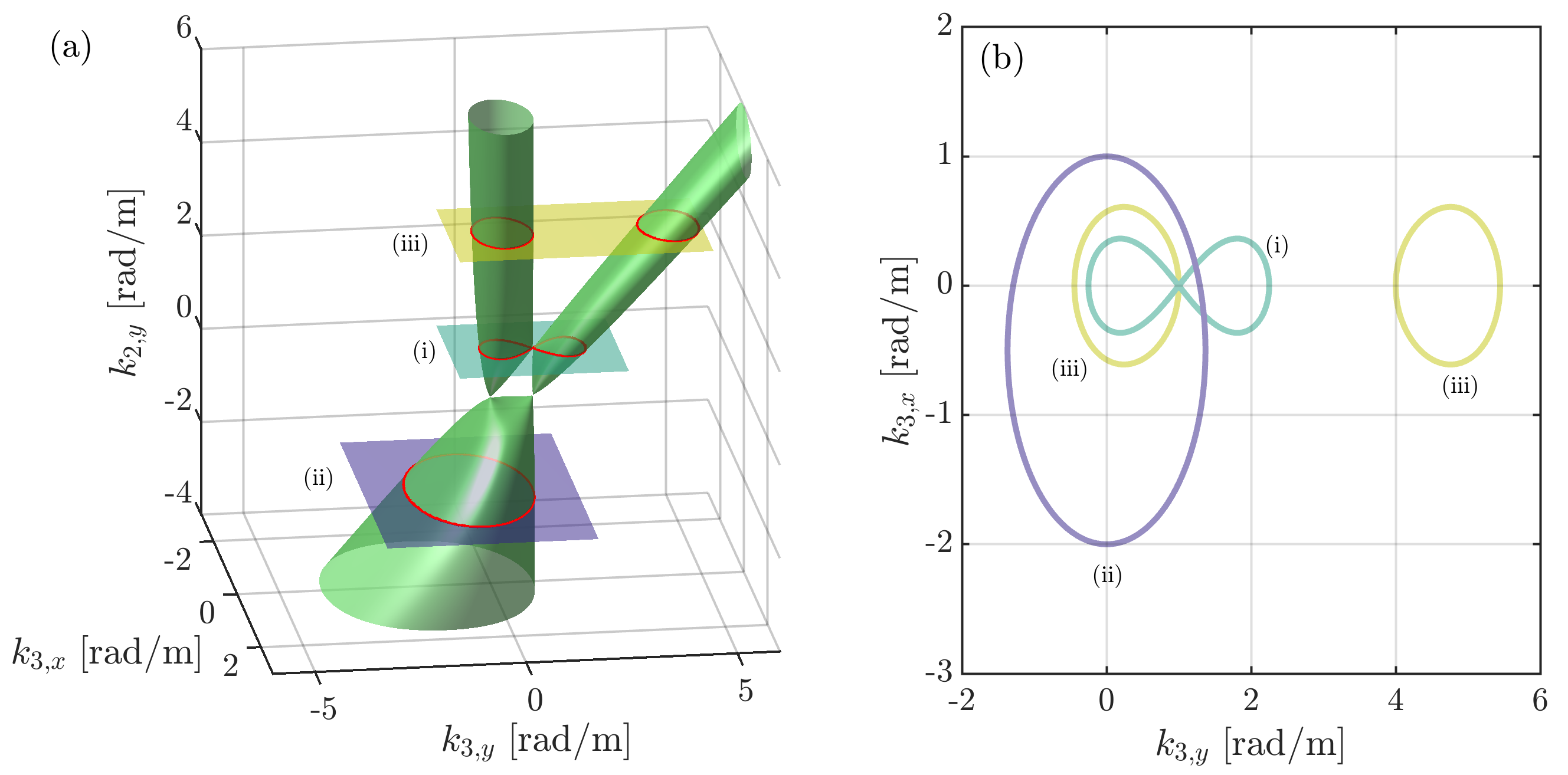}
\caption{{\bf Theoretical resonant manifold of the surface gravity wave problem.} (a) Resonant manifold (light green) intersected by three planes at different values of $k_{2,y}$, each producing a distinct intersection curve (in red): (i) the famous Phillips' figure-eight curve; (ii) a circular-like curve; (iii) a pair of closed loops. (b) 2D plot of the intersection curves extracted from the figure on the left: curve colours match the corresponding plane colours on the left.}
\label{fig:fig_1}
\end{figure}

Ocean surface gravity waves are generated by the wind. As the wind starts to blow, small ripples of just a few centimetres in wavelength form on the ocean surface. 
As they propagate under wind forcing conditions, these waves grow in length, reaching scales of hundreds of meters \cite{komen1996dynamics,janssen2004interaction}. 
From a mathematical perspective, the dynamics of ocean waves is governed by the classical equations of fluid mechanics, which are inherently nonlinear and dispersive. These two properties play a crucial role in the transfer of energy within the wave field. Starting from the inviscid water wave equations, O. Phillips found that nonlinear interactions are essential for the transfer of energy among different modes, understanding that the four-wave resonant interactions play a special role in the long term wave dynamics \cite{phillips1960dynamics}. Specifically, if one considers four wave vectors, ${\bf k}_1,{\bf k}_2,{\bf k}_3,{\bf k}_4$, the transfer of energy occurs when the following two conditions are simultaneously satisfied:
\begin{align} \label{res}
{\bf k}_1+{\bf k}_2={\bf k}_3+{\bf k}_4,\qquad   \omega_{\mathbf{k}_1}+\omega_{\mathbf{k}_2}=\omega_{\mathbf{k}_3}+\omega_{\mathbf{k}_4},
\end{align}
where $\omega_{\mathbf{k}}=\sqrt{g|\mathbf{k}|\tanh(|\mathbf{k}| h)|}$ is the angular frequency  and $g$ is the gravitational acceleration, and $h$ is the water depth. {\color{black} In deep water, the case here analyzed, it reduces to $\omega_{\mathbf{k}}=\sqrt{g|\mathbf{k}|}$}. 
  The first equation in \eqref{res} is related to the presence of cubic nonlinear terms in the equations of motion, expressed in Fourier space; this condition is also associated with the conservation of momentum in the four-wave scattering process.  The second equation in \eqref{res} represents the necessary condition for the linear growth in time of a fourth wave, given that three waves have already been excited. {\color{black}Note that the Euler equations of surface gravity, besides cubic nonlinearity (four-wave interactions), contain also quadratic nonlinearity (three-wave interaction). 
  However, in the weakly nonlinear regime in which
wave frequencies can be approximated by their linear values, those interactions are nonresonant.  This has several important
consequences (see \cite{janssen2009some}), among which is the fact that such interactions do not lead, in
a statistical sense, to an irreversible transfer of energy among modes. By using a quasi-identity canonical transformation that removes the quadratic nonresonant terms, one obtains effective equations with cubic nonlinearity that can now be resonant (four-wave interactions) \cite{zakharov2012kolmogorov,krasitskii1994reduced}.} Among all possible choices of $\mathbf{k}_1, \mathbf{k}_2$, $\mathbf{k}_3$ and $\mathbf{k}_4$, Phillips focused on those involving two equal wave vectors ${\bf k}_1={\bf k}_2$, showing that it is possible to find the other wave vectors for which Eqs.\eqref{res} are satisfied. The result is the well-known Phillips' figure-eight which can be visualized in the $(k_{3,x},k_{3,y})$-plane (Fig. 4 in \cite{phillips1960dynamics}). If $\mathbf{k}_1\neq\mathbf{k}_2$ one can still find $\mathbf{k}_3$ satisfying Eqs.\eqref{res}, leading to a series of curves in the $(k_{3,x},k_{3,y})$-plane. If we fix the values for $\mathbf{k}_1$ and $k_{2,x}$, we have a hypersurface, the resonant manifold, of which the 2D curves previously described are level sets obtained by further fixing the value of $k_{2,y}$. An example of this manifold is shown in Fig.~\ref{fig:fig_1}(a), setting $\mathbf{k}_1=(0,1)$ $\textrm{rad/m}$ and $k_{2,x}=0$ $\textrm{rad/m}$. Three examples of level curves are shown in Fig.~\ref{fig:fig_1}(b), where we recognize (i) the figure-eight curve, (ii) a circle-like curve, and a pair of closed loops (iii). 

Phillips' theoretical work, verified experimentally in wave-tank laboratories, \cite{longuet1966experiment,mcgoldrick1966measurements,waseda2015third,bonnefoy2016observation}, is deterministic and does not account for the randomness of the ocean surface. The theory of random waves was independently developed a few years later by K. Hasselmann \cite{hasselmann1962non} and V. Zakharov \cite{zakharov1966energy}. 
Starting from the inviscid equations of motion and assuming that wave amplitudes and phases are stochastic variables, they derived the so-called Wave Kinetic Equation (WKE), an integro-differential evolution equation for the wave action spectral density {\color{black}
$n(\mathbf{k},t)$}. Wave action is defined as the ratio between energy and frequency.
{\color{black}In its spatially homogeneous form, the WKE has the following form \cite{hasselmann1962non,zakharov1966energy}:

\begin{equation}\label{WKE}
    \frac{\partial n(\mathbf{k})}{\partial t} = S_{\rm nl}\,,\qquad S_{\rm nl}=\int_{-\infty}^{\infty} d\mathbf{k_1}d\mathbf{k_2}d\mathbf{k_3}|T_{\mathbf{k},\mathbf{k}_3}^{\mathbf{k_1},\mathbf{k}_2}|^2f(\mathbf{k},\mathbf{k}_1,\mathbf{k}_2,\mathbf{k}_3)\delta({\Delta_K})\delta({\Delta_{\omega}}),
\end{equation}
where $\delta$ denotes the Dirac delta, $T_{\mathbf{k},\mathbf{k}_3}^{\mathbf{k_1},\mathbf{k}_2}$ is the interaction coefficient of the four-wave interactions, and
$$f(\mathbf{k},\mathbf{k}_1,\mathbf{k}_2,\mathbf{k}_3)=n(\mathbf{k})n(\mathbf{k}_1)n(\mathbf{k}_2)n(\mathbf{k}_3)\left( \frac{1}{n(\mathbf{k})}+\frac{1}{n(\mathbf{k}_3)}-\frac{1}{n(\mathbf{k}_1)}-\frac{1}{n(\mathbf{k}_2)}\right),$$
$$\Delta_{\mathbf{k}}=\mathbf{k}+\mathbf{k}_3-\mathbf{k}_1-\mathbf{k}_2,\qquad \Delta_{\omega}=\omega_{\mathbf{k}}+\omega_{\mathbf{k}_3}-\omega_{\mathbf{k}_1}-\omega_{\mathbf{k}_2}.$$


The integrand in $S_{\rm nl}$ is the result of a theoretical closure obtained by approximating the fourth-order correlator of the wave field, the statistical object that contains all of the information about the wave-wave resonant interactions. Evaluating this correlator experimentally will be the key objective of our analysis.}
 
The WKE is analogous to the Boltzmann equation for particles: instead of two particles colliding into two particles with different momenta, it describes two waves scattering into two new waves. 
The WKE predicts that irreversible spectral transfers of wave action and energy only occurs when both conditions \eqref{res} are satisfied, thereby generalizing Phillips' result to the presence of randomness. Therefore, four-wave resonant interactions in ocean wind waves are believed to be the primary mechanism for transferring energy and wave action among scales. Today, the integral in the WKE is a fundamental element of the more general energy radiation-balance equation {\cite{komen1996dynamics}}:

{\color{black}\begin{equation}
    \frac{\partial n}{\partial t} + (\mathbf{c}_g+\mathbf{U})\cdot \nabla n\ = S_{\rm nl} + S_F + S_D\,,
\end{equation}

This equation also includes nonhomogeneous transport terms such as linear dispersive propagation with the group velocity $\mathbf{c}_g$ and advection by the mean current $\mathbf{U}$ on the left-hand side, and the so-called source terms on the right-hand side such as nonlinear wave-wave interactions $S_{\rm nl}$, nonlinear white-capping dissipation $S_D$ and wind forcing $S_F$. While $S_{\rm nl}$, the same term as in Eq.~\eqref{WKE}, is derived from primitive equations, $S_D$ and $S_F$ are the fruit of decades-worth of parameterization efforts~\cite{janssen2004interaction}.}
The complete model serves as the foundation for operational wave forecasting systems \cite{cavaleri2007wave,komen1996dynamics,janssen2004interaction}, which are coupled with atmosphere and, often, with ocean general circulation models. Those models run daily to predict the statistical properties of the ocean surface worldwide. {\color{black}Note that we have omitted further coupling terms such as shear currents, bottom friction and wave-bottom scattering.}

{\color{black}Despite the scientific significance of the theory and the impact of its applications, and despite the fact that there is indirect evidence of the existence of four-wave resonant interactions, a direct investigation of the correlators related to four-wave resonant interactions in ocean wind waves is still lacking}. 
Historically, measurements have been conducted using buoys, which record surface elevation over time but lack spatial information. Today, with stereoscopic or lidar measurements \cite{benetazzo2006measurements,benetazzo2012offshore, AirborneObservationsofFetchLimitedWavesintheGulfofTehuantepec,lenain2017measurements}, it is possible to accurately capture the surface elevation $\eta(x,y,t)$  both in time $t$ and space (the two horizontal dimensions $x$ and $y$) over regions spanning thousands of square meters. This marks a significant advance in the investigation of resonances in the ocean.

Features supporting resonant interactions in the ocean already exist. For example, the WKE predicts out-of-equilibrium stationary solutions characterized by power laws for the energy and wave action spectra, i.e. the two Kolmogorov-Zakharov solutions with constant energy flux (direct cascade) and constant wave action flux (inverse cascade) in Fourier space \cite{zakharov1966energy,zakharov2012kolmogorov}. 
For surface gravity waves, theoretical predictions suggest that, {\color{black} in the idealized case of a large scale forcing and a small scale dissipation}, the {\color{black} direct cascade}  frequency spectrum scales as $\omega^{-4}$, while the wave number spectrum follows $ |\mathbf{k}|^{-2.5}$. {\color{black}On the other hand, when the forcing is at small scale and damping at large scale the inverse cascade develops and it is characterized by frequency spectrum scaling as $\omega^{-11/3}$ and the wave number spectrum scaling as $|\mathbf{k}|^{-7/3}$. 
} 
Spectral slopes consistent with the prediction of a direct energy cascade have been observed in the ocean 
\cite{kawai1977field, hwang2000airborne, resio2004equilibrium}.
A second indication of resonant interactions in the ocean is the formation of a bimodal spectrum at high frequencies observed experimentally \cite{ewans1998observations,romero2010airborne,lenain2017measurements}. This feature is also detected from numerical simulations of the WKE \cite{ badulin2017ocean} or deterministic simulations of the Euler equations, \cite{toffoli2010development}. 

Despite these two observations are compatible with nonlinear interaction mechanisms, they represent only an indirect proof of the existence of resonances; indeed, it could be argued that power-law spectra and bimodality could be the result of dissipation and forcing. {\color{black}For instance, Phillips~\cite{phillips1985spectral} suggested that the same scaling of the direct-cascade Kolmogorov-Zakharov solution $\sim k^{-2.5}$ can be obtained as the result of a balance between $S_{\rm nl}$, $S_D$ and $S_F$, if the latter two are parameterized so that all three terms are proportional to each other and of similar order of magnitude. This gives rise to a more nuanced interpretation of the $\sim k^{-2.5}$ range in which nonlinear interactions are an important piece but not necessarily the dominant one. As shown in \cite{romero2012spectral,lenain2017measurements}, high nonlinearity is today considered responsible for a transition to the steeper saturation spectrum $\sim k^{-3}$ \cite{phillips1958equilibrium} at high wave numbers (where the transition wave number depends on the level of nonlinearity in a given wave field).
Regarding bimodality, it has also been  suggested that it may be the result of long-short wave interactions or of dissipation that is larger in the direction of the long dominant waves \cite{romero2019distribution,akaawase2025observations}.}

{\color{black}Here, we focus on identifying a methodology to detect resonant interactions in the ocean.  In view of the various possible interpretations of the observations, we wish to  consider an observable that uniquely characterizes the four-wave resonant interactions.} Therefore,  based on theoretical considerations on the WKE, our strategy consists of computing the spectral fourth-order correlator directly from oceanic experimental data~\cite{campagne2019identifying}, and thereby give direct evidence of the existence of  resonances or quasi-resonances (i.e. imperfect resonances due to nonlinear broadening of the exact resonant condition) in real-ocean wind waves.

\section*{Results}
\subsection*{Theory of resonant wave interaction}\label{sec2}
To properly grasp the data analysis presented in the next section and the notation we use, we find it helpful to first provide a brief review of the theory of resonant interactions \cite{phillips1960dynamics,hasselmann1962non,Zakharov1967,benney1962non,benney1966nonlinear,benney1969random,longuet1962resonant,zakharov1966energy}. 
We follow the Hamiltonian formulation of Zakharov \cite{Zakharov1967,krasitskii1994reduced}, in which the surface elevation $\eta(x,y,t)$ and the surface velocity potential $\psi(x,y,t)$ form a pair of canonically conjugated variables. The theoretical derivation needed for our work, reviewed  in {\it Methods}, is made of two parts: the first one includes a series of analytical manipulations at the deterministic level, while the second one consists of taking suitable statistical assumptions for the random wave field and suitable limits on the time and space scales.

The first part includes the following steps. First, by expanding the velocity potential about the surface at equilibrium, it is possible to write the equations of motion as a system of coupled evolution equations in Fourier space. Second, we introduce the normal variable $a_{\mathbf{k}}(t)$, a linear combination of the Fourier transform of both the surface elevation and the surface velocity potential \cite{krasitskii1994reduced},
\begin{equation}\label{ak}
a_{\mathbf{k}}(t)=\sqrt{\frac{g}{2\omega_{\mathbf{k}}}}\eta_{\mathbf{k}}(t)+
i \sqrt{\frac{\omega_{\mathbf{k}}}{2g}}\psi_{\mathbf{k}}(t).\qquad 
\end{equation}
Third, a near-identity canonical transformation is performed, defining a new variable $b_{\mathbf{k}}(t)$ \cite{janssen2009some,krasitskii1994reduced}. From a physical point of view, this transformation eliminates the so-called bound modes, i.e. those modes that do not oscillate with the linear dispersion relation. For example, in a Stokes expansion with fundamental mode ${\bf k}=(k_0,0)$, the first harmonic $2k_0$ does not obey the linear dispersion relation $\omega(2 k_0)$,  and it travels with the same phase velocity as the fundamental mode. The mode $2k_0$ is bound to $k_0$ and it will oscillate with frequency  $2\omega(k_0)$. Eliminating bound modes in a random sea state is essential to isolate resonant interactions. The evolution of the new variable $b_{\mathbf{k}}(t)$ is given by the so-called Zakharov equation.

In the second part of the derivation, since we are interested in a statistical theory of waves, we assume that $b_{\bf k}=J_{\bf k}e^{i\theta_{\bf k}}$ are random variables with {\color{black}independent amplitudes $J_{\bf k}$ and independent and identically distributed phases $\theta_{\bf k}$;} moreover, phases are uniformly distributed in the interval $[0,2\pi)$.
This prescribed statistics defines our ensemble of random data, and we denote the average of an observable over these random data (ensemble average) by $\langle\dots\rangle$. Following \cite{janssen2004interaction}, it is possible to derive an evolution equation for the lowest-order nontrivial observable, the wave-action spectral density $\langle|b_{\mathbf{k}}|^2\rangle$, which depends on the imaginary part of the fourth-order spectral correlator $\langle b_{\mathbf{k}_1}^*(t)b_{\mathbf{k}_2}^*(t)b_{\mathbf{k}_3}(t)b_{\mathbf{k}_4}(t)\rangle$. In the large-time and large-box limits, this correlator becomes proportional to the product of two delta functions involving four wave vectors and the corresponding frequencies \cite{annenkov2018}, (see  {\it Methods}):
\begin{align}\label{eq:6}
\mathfrak{Im}[\langle b_{\mathbf{k}_1}(t)b_{\mathbf{k}_2}(t)b_{\mathbf{k}_3}^*(t)b^*_{\mathbf{k}_4}(t)\rangle]\sim\delta\left(\omega_{\mathbf{k}_1}+\omega_{\mathbf{k}_2}-\omega_{\mathbf{k}_3}-\omega_{\mathbf{k}_4}\right)\delta\left(\mathbf{k}_1+\mathbf{k}_2-\mathbf{k}_3-\mathbf{k}_4\right),
\end{align}
where $\mathfrak{Im}[...]$ is the imaginary part and $\delta$ is the Dirac delta (strictly speaking, the last relation is valid only when the deltas are considered under integral signs). The appearance of the two Dirac deltas in the above expression plays a role analogous to the constraints given in Eq.\;\eqref{res}, highlighting that these interactions are the only ones that on average survive in the evolution of a random wave field. The calculation just outlined leads to the WKE~\eqref{WKE}, which yields a nonzero transfer of energy among the Fourier modes only when both delta function conditions are satisfied. {\color{black}In order to reach the closed form~\eqref{WKE} from terms of the type~\eqref{eq:6}, a further expansion to higher order and random-phase averaging are needed, as outlined in {\it Methods}.}

\subsection*{Dataset and data analysis}\label{dataset}
\begin{figure}
    \centering
\includegraphics[width=\textwidth]{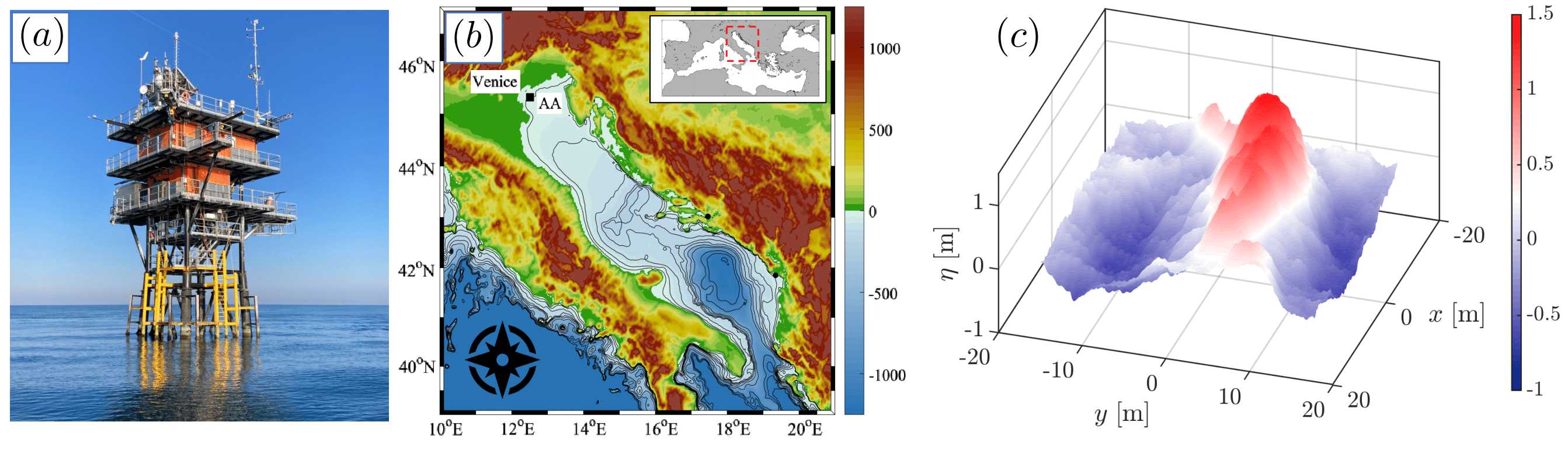}
  \caption{{\bf Sea surface elevation measurements from the Acqua Alta oceanographic tower.} (a) The Acqua Alta oceanographic tower managed by CNR-ISMAR. (b)  Geographical position of the tower in the Adriatic sea outside of the Venice lagoon, with the color indicating bathimetric elevation in meters with respect to mean sea level. The inset shows a map of the Mediterranean sea, with the dashed red box outlining the boundaries of the region shown in panel ~\ref{fig:fig_2}(b).  (c) Snapshot of measured surface elevation $\eta$ in the squared field of side $32.6$ m used in our analysis, as reconstructed by stereographic technique. The colorbar indicates sea surface elevation in meters with respect to mean sea level.}
  \label{fig:fig_2}
\end{figure}

The dataset employed in this study was sourced from the public data repository documented in \cite{Guimaraes2020}. Original data were collected from the ``Acqua Alta" oceanographic research tower (see Fig.~\ref{fig:fig_2}(a)), situated in the northern Adriatic Sea (Italy) at coordinates $ 45.32^\circ $ N, $ 12.51^\circ $ E. The tower is located 15 km offshore from the Venice littoral at a water depth of $h=17$ m (see Fig.~\ref{fig:fig_2}(b)). The data were acquired using two stereo cameras capturing the same sea surface area. The wave elevation $\eta(x,y,t)$ observation was obtained using a stereographic technique \cite{benetazzo2006measurements,bergamasco2017wass}, with examples of measured wave field shown in Fig.~\ref{fig:fig_2}(c).
{\color{black}The dataset here considered was collected on 27 March 2014, starting the acquisition at 09:10 UTC, and it comprises about 28 minutes of surface elevation recordings over a squared area of side $\ell=32.6$ m, with a sampling rate of 12 Hz. Regarding in-situ local observation, during the recording period the wind parameters were measured by a VT0705B SIAP anemometer and the local wave parameters by a Nortek Acoustic Wave and Current pro ler, AWAC.  The emerging picture sees a local northeasterly wind (with mean direction of
56° N) with mean speed of 10.6 m/s (at 17.5 m height).
Locally wind generated waves had limited fetch, total signicant wave height $H_{s}$ = 1.38 m, peak period
$T_{p}$ = 4.9 s, and peak direction of wave propagation $\theta_{p}$ = 257 N. 
The sea current, measured by the same instrument AWAC, was almost uniform along the vertical direction (from the surface to the sea bottom), with a mean speed of 0.25 m/s and mean direction of propagation around 230° N. Note that  for the longest waves lengths resolved in the experimental field of about 30 m in 17 m-deep water, the dispersion relation can be approximated to the one in infinite water depth, with an error on the frequency smaller than 0.1$\%$. For all our purposes, we will therefore assume the deep-water dispersion relation in the following.}

To compute the fourth-order correlator, one first needs to build the variable $a_{\mathbf k}$ as expressed Eq.\;\eqref{ak}. With our experimental apparatus, a direct measurement of the velocity potential is not possible; therefore, we use the leading-order relation: 
\begin{equation}
\psi_{\mathbf{k}}(t)=g\frac{\dot{{\eta}}_{\mathbf{k}}(t)}{\omega^2_{\mathbf{k}}},
\end{equation}
where $\dot{{\eta}}_{\mathbf k}(t)$ is the time derivative of ${\eta}_{\mathbf k}(t)$.
Moreover, it is essential to apply a filtering procedure to the surface elevation. Due to the system's nonlinearity, waves at a given wave vector not only oscillate at the frequency prescribed by the dispersion relation, but can also become locked to other waves through a nonlinear process. {\color{black}In weakly nonlinear conditions, it is the quadratic nonlinearities, which are nonresonant, that are responsible for the generation of these bound modes \cite{krasitskii1994reduced,janssen2009some}. They have important effects on the wave field, such as introducing a skewness in the solution of the wave equations leading to rounder troughs and sharper crests. However, it is important to stress that the bound modes are dynamically independent from the free modes that participate in the resonant four-wave interactions.}

Since resonances occur for waves oscillating according to the dispersion relation, a filtering procedure to remove bound modes becomes crucial, and is the empirical analogue of constructing the variable $b_{\mathbf{k}}\left(t\right)$ introduced above. The filtering procedure (see {\it Methods}) 
yields the filtered surface elevation $\hat\eta(x,y,t)$. 
\begin{figure}
\centering
\begin{minipage}{\linewidth}
\centering
\includegraphics[width=\linewidth]{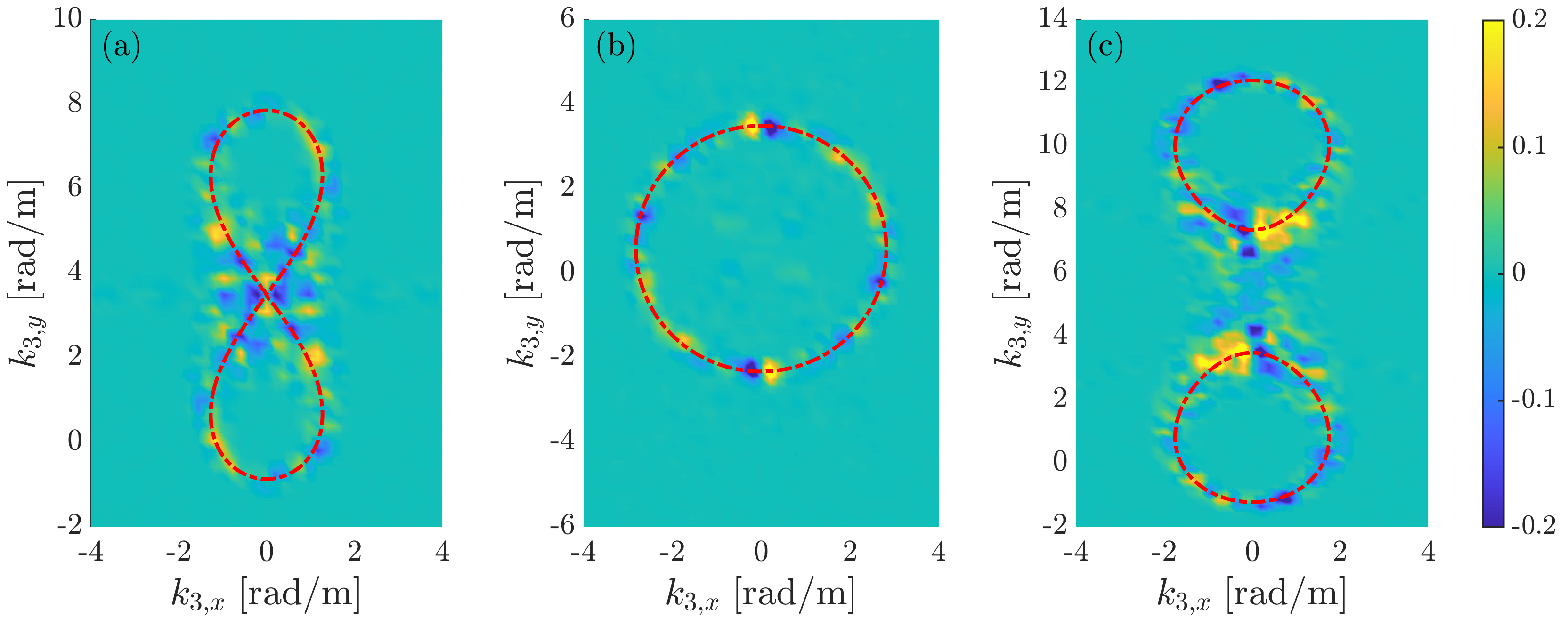}
\caption{{\bf Observationally-reconstructed sections of the resonant manifold.} Numerical evaluation of the imaginary part of the normalized four-point correlator $C(\mathbf{k}_1,\mathbf{k}_2,\mathbf{k}_3)$ defined in Eq.\;\eqref{c2} shown as a function of $\mathbf{k}_3$. The values of $\mathbf{k}_1$ and $\mathbf{k}_2$ for the three cases are (a) $\mathbf{k}_1=\mathbf{k}_2=(0,3.42)$ ${\rm rad/m}$, (b) $\mathbf{k}_1=(0,3.42)$ ${\rm rad/m}$ and $\mathbf{k}_2=(0,-2.28)$, (c) $\mathbf{k}_1=(0,3.42)$ ${\rm rad/m}$ and $\mathbf{k}_2=(0,7.22)$ ${\rm rad/m}$. The red dashed lines represent the theoretical curves, analogous to those shown in Fig.\;\eqref{fig:fig_1}(b). The colorbar ranges from large positive values (yellow) to large negative values (blue) of the four-point correlator. Large absolute values of the correlator are observed along the theoretical curves, and negligible values are found elsewhere.}
\label{fig:fig_3}
\end{minipage}
\end{figure}
The quantity of interest here is the four-wave correlator $\langle b_{{\bf k}_1}(t)b_{{\bf k}_2}(t)b_{{\bf k}_3}^*(t) b_{{\bf k}_4}^*(t)\rangle$, computed for wave vectors that satisfy momentum conservation ${\bf k}_4={\bf k}_1+{\bf k}_2-{\bf k}_3$. According to Eq.\;\eqref{eq:6}, in the long-time and large-box limits, the correlator should only have contributions on the resonant manifold. 
Therefore, we focus our analysis  on the evaluation of the normalized four-point correlator defined as (see \cite{campagne2019identifying,Zhang2022}
)
\begin{align}\label{c2}
C(\mathbf{k}_1,\mathbf{k}_2,\mathbf{k}_3)=\frac{\langle b_{\mathbf{k}_1}(t)b_{\mathbf{k}_2}(t)b^*_{\mathbf{k}_3}(t)b^*_{{\bf k}_1+{\bf k}_2-{\bf k}_3}(t)\rangle}{\langle|b_{\mathbf{k}_1}(t)|\rangle\langle|b_{\mathbf{k}_2}(t)|\rangle\langle|b^*_{\mathbf{k}_3}(t)|\rangle\langle|b^*_{{\bf k}_1+{\bf k}_2-{\bf k}_3}(t)|\rangle}
.
\end{align}
To reproduce the  Phillips' eight-figure, we compute the imaginary part of the above quantity fixing two wavevectors to be equal ($\mathbf{k}_1=\mathbf{k}_2$), and imposing the condition ${\bf k}_4={\bf k}_1+{\bf k}_2-{\bf k}_3$. In the specific case, we choose $\mathbf{k}_1=\mathbf{k}_2=(0,3.42)$ ${\rm rad/m}$. The result is shown in Fig. \eqref{fig:fig_3}(a). We also investigate different configurations: by setting $\mathbf{k}_1=(0,3.42)$ ${\rm rad/m}$ and $\mathbf{k}_2=(0,-2.28)$, we find the circular-like curve shown in Fig.\;\eqref{fig:fig_3}(b),  and finally, setting $\mathbf{k}_1=(0,3.42)$ ${\rm rad/m}$ and $\mathbf{k}_2=(0,7.22)$ ${\rm rad/m}$, we find the pair of closed loops shown in Fig.\;\eqref{fig:fig_3}(c). On the same figures, we have plotted the theoretical predictions. As is clear from the figures, the correlator is different from zero only in the proximity of the curves describing the exact resonances and it is zero elsewhere. We observe that along the resonant curves, both positive and negative correlations are present, shown in yellow and blue, respectively. It is worth pointing out that in Eq.\;\eqref{c2} $\langle \dots \rangle$ denotes the average over the ensemble, while in our analysis we performed an average over time. Indeed, identifying the two averaging procedures needs the assumptions of statistical stationarity  and some notion of ergodicity, both of which cannot be proven in our dataset. Despite the conceptual difference, we find remarkable agreement between our experimental results and the theoretical predictions. We repeated the analysis using different values of the filter width and observed that, when the width is small enough, the results persist; while, as the width increases, as expected, the correlator signal progressively weakens, eventually vanishing for sufficiently large values.
\begin{figure}
\centering
\begin{minipage}{\linewidth}
\centering
\includegraphics[width=\linewidth]{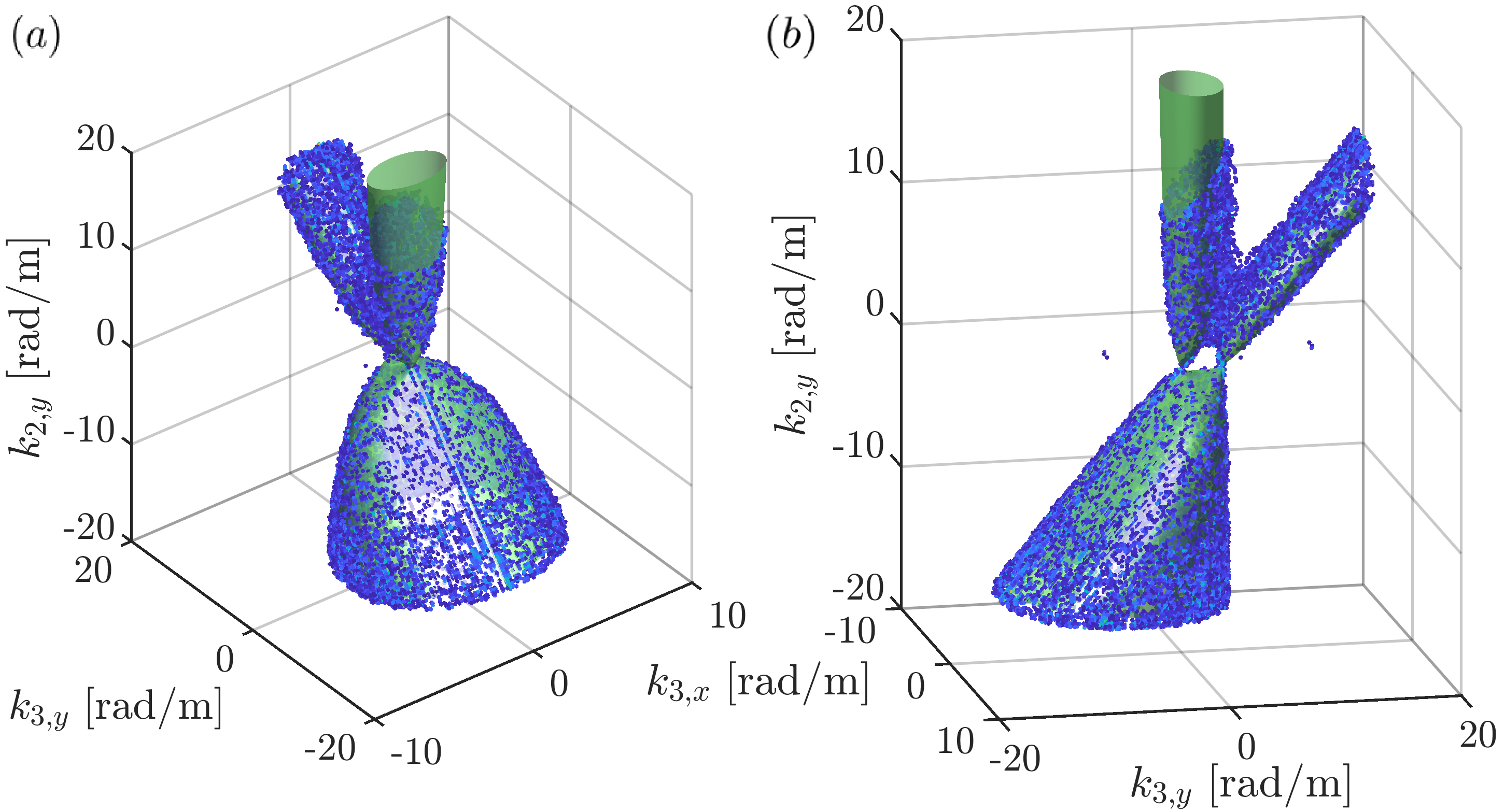}
\caption{{\bf Observational reconstruction of the full resonant manifold.} Two different views of the reconstructed three-dimensional manifold by evaluating the absolute value of the imaginary part of the normalized four-point correlator $C(\mathbf{k}_1,\mathbf{k}_2,\mathbf{k}_3)$ defined in Eq.\;\eqref{c2} as a function of $\mathbf{k}_3$ and $k_{2,y}$. Blue dots represents the points where $|\mathfrak{Im}\left\{C(\mathbf{k}_1,\mathbf{k}_2,\mathbf{k}_3)\right\}|$ is {\color{black}above-threshold} while the green surface is the theoretical manifold defined by the resonant conditions \eqref{res}. For this particular configuration we have $\mathbf{k}_1=(0,3.42)$ ${\rm rad/m}$, and $k_{2,x}=0$ ${\rm rad/m}$. While panel (a) emphasizes the overall structure, panel (b) showcases the nontrivial hole in the resonant manifold (both theoretical and reconstructed) in the neighborhood of the origin.}
\label{fig:fig_4}
\end{minipage}
\end{figure}
The experimental three dimensional resonant manifold, similar to the theoretical one shown in Fig.\;\eqref{fig:fig_1}, is now evaluated by computing the normalized four-point correlator \eqref{c2} by fixing $\mathbf{k}_1=(0,3.42)$ ${\rm rad/m}$ and $k_{2,x}=0$ and averaging over time. In this part of the analysis, we are no longer interested in the sign of the correlation, but rather in identifying the regions where the correlation is non-zero. To this end, we compute the absolute value of the imaginary part of the correlator and the result is shown in Fig.\;$\eqref{fig:fig_4}$. Since this quantity 
is non-zero only on the resonant manifold, we plot points where it exceeds a threshold of 0.08. The experimental three dimensional resonant manifolds are presented in Fig. \eqref{fig:fig_4} from two different points of view. 
The plot reveals that the points cluster near the resonant surface. {\color{black} We show in the {\it Methods} section that all wave numbers involved in the resonance sets shown in Figs.~\ref{fig:fig_3}-\ref{fig:fig_4} appear to be in the main $\sim k^{-2.5}$ range of the spectrum.}




\section*{Discussion}\label{sec3}
Predicting the spectral distribution of energy in oceanic surface wind waves is a difficult task for numerical models. 
It is understood that four-wave resonant interactions play a fundamental role in transferring energy through the wave scales: inclusion of wave-wave interactions (even if simplified, see \cite{komen1996dynamics}) in the radiation-balance equation allows the models to do a fairly good job at predicting sea states. 
Indeed, the reliability of such predictions is essential, for example, for safe navigation across the global seas. However, because of the difficulty of measuring the evolution of spatially extended portions of sea surface at high space and time resolutions, the presence of resonant wave interactions in oceanic wind wave fields has never been confirmed directly. Therefore, validating with field data the wave-wave interaction theory underlying the model parameterisations remains a high priority. The obvious repercussions involve, among others, increasing confidence in the models and, where necessary, improving their accuracy and performance.

Here, we used long high-resolution time series of sea surface elevation reconstructed by stereographic cameras in an observational field of about 1000 square meters. The high resolution and large width of the resulting space-time domain have allowed us to build the spectral fourth-order correlators of the wave field. These correlators are those that {\color{black}are modeled in} the WKE governing the nonlinear spectral transfers among waves of different wavelengths. {\color{black}However, the WKE is based on some statistical assumptions about these wave field correlators, which so far have not been directly experimentally validated in the field.}

In a random linear wave field, waves of different wavelengths would be uncorrelated. When nonlinearity is at play, it creates correlations in two different ways~\cite{onorato2013rogue}. The first one is through bound modes, as any fundamental ``free'' mode of oscillation is accompanied by the suite of its higher-order harmonics (called the Stokes series). 
The second is through actual nonlinear interaction, resonant or non-resonant, between the free modes of the system. {\color{black} In weakly nonlinear wave fields, the two mechanisms are independent of each other and can be disentangled.} In our analysis, we have eliminated the bound modes by applying a spectral filter, ensuring that any observed correlation must be due to nonlinear interaction between free modes.

When plotting the fourth-order correlators between four independent wave vectors, we observed significant values only in close proximity to a hyper-surface in the high-dimensional space spanned by the four wave vectors. Such empirically reconstructed hyper-surface exactly reproduces the theoretical resonant manifold predicted by the resonant theory of four-wave interactions. 
By taking several lower-dimensional sections of the hyper-surface, we have shown that the empirically reconstructed resonant manifold matches the theoretical one, with the maximum value of the fourth-order correlator in correspondence of the exact resonances, and also including a broadened finite-width contribution from near resonances. 
In particular, we have observed the renowned Phillips' figure-eight resonances, predicted almost seven decades ago for the particular case with two equal wave vectors. Furthermore, we have shown the existence of many more resonances with qualitatively different patterns, all closely corresponding to Hasselmann's theory of four-wave interactions. The analysis shows beyond doubt that all types of four-wave resonant and near-resonant interactions predicted by the theory are strongly active in the observed surface wave field. {\color{black} Furthermore, this demonstrates that, within our approach, it is possible to study  correlation of arbitrary order directly from the field, provided their convergence.}

Our results pave the way for many more experiments in which all aspects of wave-wave interaction theory can be investigated {\color{black}with direct ground-truthing against field data}. Such experiments should be designed at locations with different statistics of sea states to test the details of wave-wave resonant theory under different forcing conditions. Here, we have established a prototype procedure for the direct exploration of wave-wave interactions in surface wave fields that will represent a fundamental step toward validating and improving models for the prediction of wind wave fields.
{\color{black} Obviously, the wave motions in the ocean are far more complex than the existing models, where many factors are ignored; the effects of
such a neglect are often difficult to evaluate. The direct approach pioneered in this
work can clarify a number of old open questions.}

\section*{Methods}\label{methods}

\subsection*{Processing data from Acqua Alta Oceanographic Tower}

\begin{figure}
\centering  \includegraphics[width=1\linewidth]{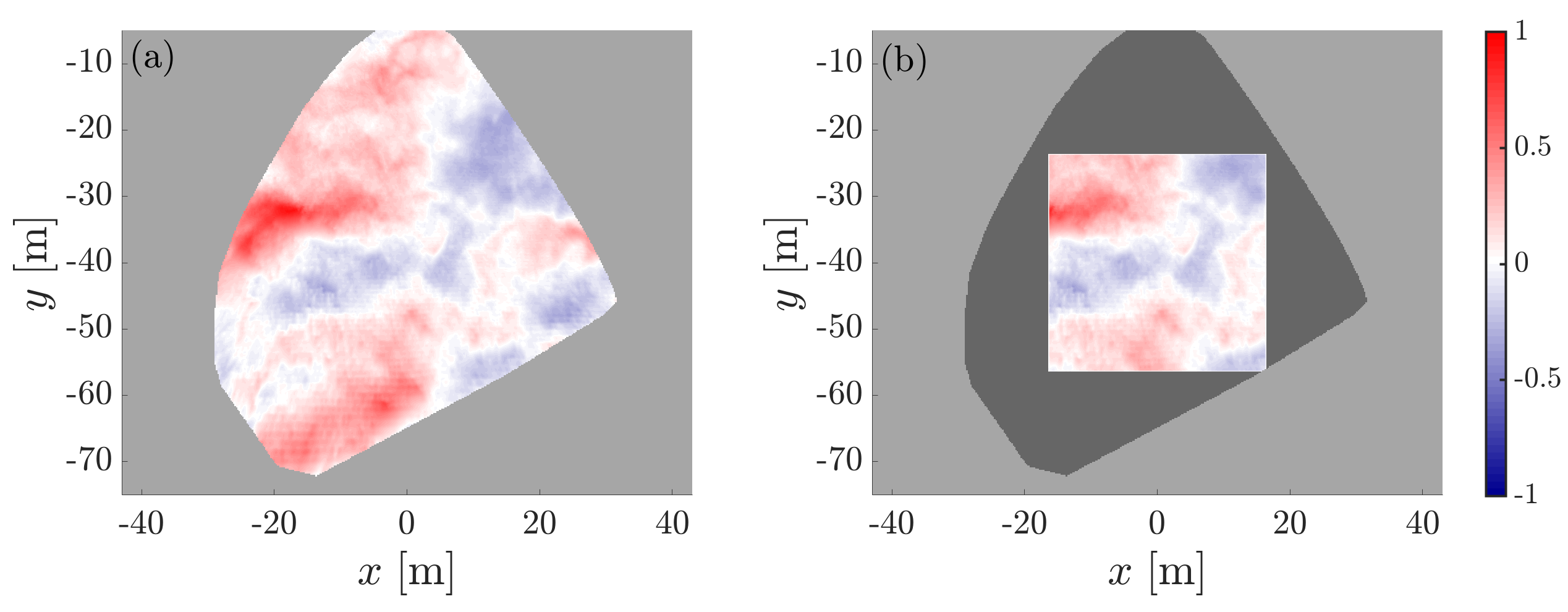}
\caption{{\bf Stereographic processing of the sea surface elevation.} (a) The rectangular region of the sea recorded by the cameras, and the successfully reconstructed sea surface (coloured region) after applying the mask. The light grey area represents burnt data points. (b) The reduced area used in our data analysis: the dark grey region represents the additional removed data points. In both panels, the colorbar indicates the sea surface elevation in meters, ranging from deep troughts in dark blue to tall crests in dark red.}
\label{fig:fig_5}
\end{figure}

Data are stored in a NetCDF-4 format (.nc file) containing metadata, such as the acquisition frame rate, the spatial resolution, and the distance of the cameras from the sea level. 
In addition, the file includes two bidimensional grids providing spatial information for the coordinates, a time array storing the acquisition times, a three-dimensional array representing the elevation of the surface $\eta(x,y,t)$, and a mask containing information, stored as {\it NaN}, about the points where the reconstruction algorithm failed. In our analysis, we considered the file \textit{Surfaces\_20140327\_091000.nc} which contains the data recorded on March 27, 2014, at 9:10:00 AM. The dataset covers a rectangular region with sides $L_x=86$ and $L_y=70$ m, with a spatial resolution of $\Delta x=\Delta y=0.2$ m.  The time series consists of $20,000$ samples acquired at a sampling frequency of $f_s=12$ Hz, corresponding to a total duration of approximately 28 minutes. The water depth is $h=17$ m and hence, the dispersion relation used in all calculations is $\omega_{\mathbf{k}}=\sqrt{g|\mathbf{k}|}$. In Fig.\;\eqref{fig:fig_5}(a) we show one snapshot of the successfully reconstructed surface elevation after applying the mask (light grey region). Moreover, starting from the centre of the available domain, we select the largest possible square region, expanding outward until the first {\it NaN} value is found in the mask. As a result of this procedure, we obtain data corresponding to a region with side lengths of $\ell=32.6$ m yelding a meshgrid of $163\times163$ points (see Fig.\;\eqref{fig:fig_5}(b)).  Therefore, the dataset used in our analysis consists of a three-dimensional array of size $(163,163,20000)$. 
In Fourier space, the modes are uniformly spaced with a wavenumber increment $\Delta k=\Delta k_x=\Delta k_y=2\pi/\ell\approx 0.19\;\rm rad/m$.

{\color{black}\subsection*{Wave number and frequency spectra}
To characterize the statistical state of the wave field, We start by evaluating the wave number power spectrum by applying a two-dimensional Hann window along the spatial directions of the surface elevation $\eta(x,y,t)$ for all time points, in order to enforce periodicity in both directions. We first remove the mean value and then compute the two-dimensional Fourier transform for all time points, calculate its modulus squared, and average over time. Finally, we integrate over $k=|\mathbf{k}|$ from $k=0$ to $k_{\textrm{MAX}}=\sqrt{k_{x,\textrm{MAX}}^2+k_{y,\textrm{MAX}}^2}$ using $N_{\text{bins}}=100$. The result is presented in Fig.\;\eqref{fig:fig_6}(a) along with the theoretical prediction $k^{-5/2}$ (red dashed line). The  frequency power spectrum is evaluated by fixing a point in space and considering its variation over time. 
A one-dimensional Hann window is applied to enforce periodicity, and the mean value is removed from the dataset to focus on the motion about the mean of the surface elevation. The Fourier transform is then computed for this temporal signal. This procedure is repeated for the points of the surface elevation selected on a two-dimensional grid with a 1-meter distance between the points, which corresponds to five times the spatial resolution of the data. We indicate these points as $\mathcal{D}$. Finally, we average over $\mathcal{D}$. 
The result is presented in Fig.\;\eqref{fig:fig_6}(b) together with the theoretical prediction $\omega^{-4}$ (red dashed line).
\begin{figure}
    \centering  \includegraphics[width=1\linewidth]{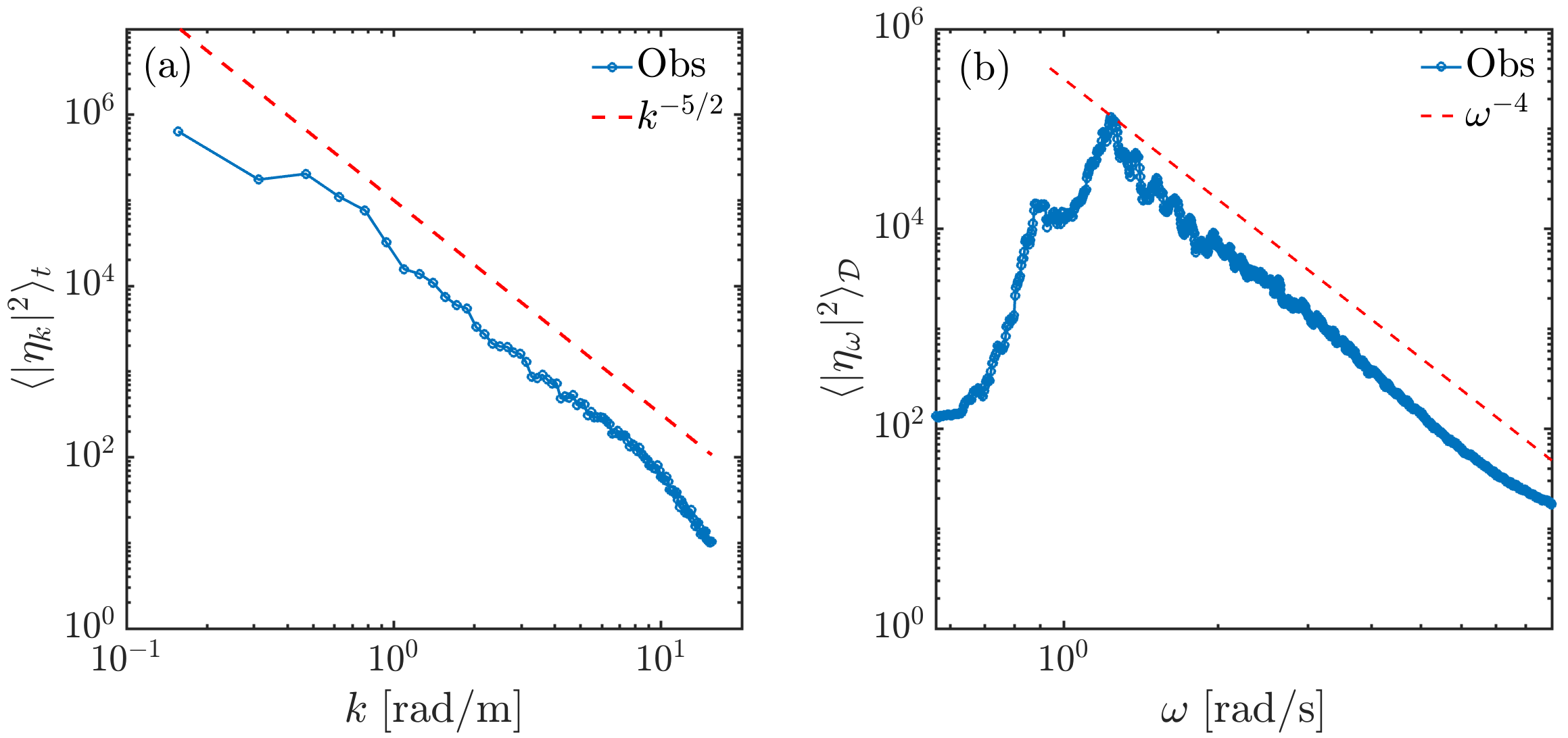}
    \caption{{\bf Omnidirectional spectra of surface gravity wave energy.} (a) The wave number power spectrum $\langle |\eta_k|^2\rangle_t $ averaged over time with the theoretical prediction (red dashed line) $k^{-5/2}$. (b) The frequency power spectrum $\langle |\eta_{\omega}|^2\rangle_{\mathcal{D}}$ averaged over the sampling domain $\mathcal{D}$ with the theoretical prediction (red dashed line) $\omega^{-4}$.}
    \label{fig:fig_6}
\end{figure}

It is known that the transition between the $k^{-2.5}$ and $k^{-3}$ scalings should scale with the friction velocity $u_*$ \cite{lenain2017measurements}. The estimate of the normalized friction velocity following the formula in \cite{young1999wind} gives $u_*/\sqrt{gH_s}\simeq0.41$, which implies a transition wave number of about $10$ rad/m \cite{lenain2017measurements}. This seems to coincide with the transition from $k^{-2.5}$ to a steeper spectral slope in Fig.~\ref{fig:fig_6}(a), although it happens close to the highest resolved wave number and therefore the range to be able to detect the transition is very limited. Regardless, an important aspect is that all wave numbers involved in the resonance sets shown in Figs.~\ref{fig:fig_3}-\ref{fig:fig_4} appear to be in the main $\sim k^{-2.5}$ range, consistently both with the Kolmogorov-Zakharov direct cascade and with Phillips' ``equilibrium range''.}

\subsection*{Evaluation of the wavenumber-frequency spectrum}
\begin{figure}
\centering\includegraphics[width=0.6\linewidth]{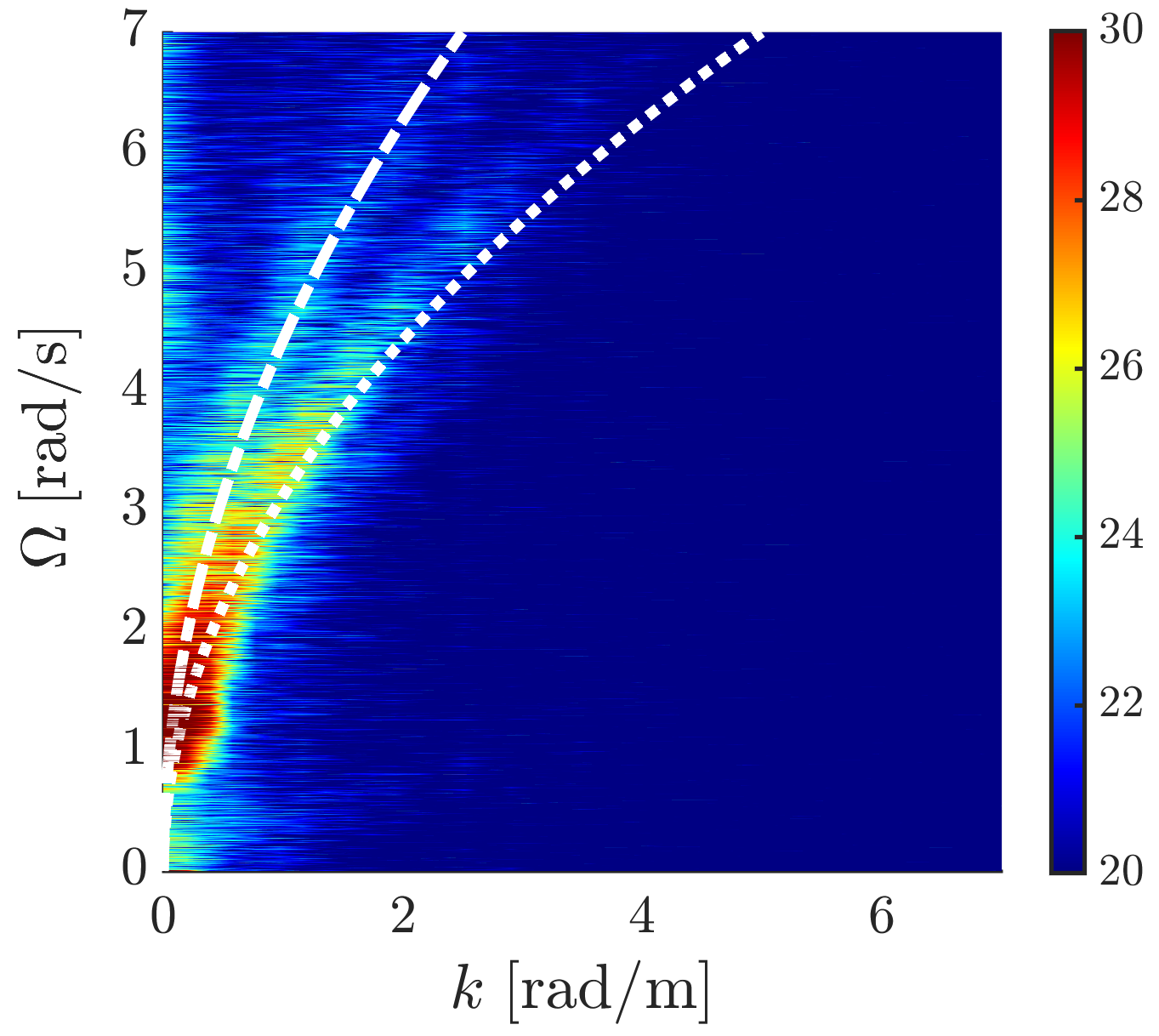}
\caption{{\bf Two-dimensional wavenumber-frequency spectrum.} The colormap shows the logarithm of the wavenumber-frequency power spectrum: the lower dashed line is $\Omega=\sqrt{g|\mathbf{k}|}$, i.e. the free-modes dispersion relation, and the upper dashed line is $\Omega=\sqrt{2g|\mathbf{k}|}$, corresponding to the first harmonics of bound modes. }
\label{fig:fig_7}
\end{figure}

To evaluate the wavenumber-frequency, ($k-\Omega$), power spectrum of our data, we first compute the two-dimensional Fourier transform $\eta_{\mathbf{k}}(t)$ along the spatial directions, then construct the auxiliary complex variable \eqref{ak} and subsequently perform the one-dimensional Fourier transform along the temporal direction. In the second expression of \eqref{ak} we first evaluate the first-order time derivative in the physical space and then compute the two-dimensional Fourier transform. Before evaluating each Fourier transform, we apply Hann windows to ensure periodicity. We finally integrate over the angles to obtain a function of $|\mathbf{k}|$. The logarithm of the ($k-\Omega$) power spectrum is presented in Fig.\;\eqref{fig:fig_7}: the lower red dashed lines represents the dispersion relation $\Omega=\sqrt{g|\mathbf{k}|}$ for surface gravity waves, i.e., it corresponds to frequency at which {\it free} waves oscillates; while the upper line $\Omega=\sqrt{2g|\mathbf{k}|}$ corresponds to those waves which do not satisfy the linear dispersion relation, and are known as {\it bound} waves. From a theoretical point of view, such waves can be eliminated from the theory using a a near identity transformation which is briefly discussed below.

\subsection*{Spectral filter to remove bound modes}

To reveal the resonant interactions, we follow \cite{campagne2019identifying,Zhang2022} and
remove the bound modes by multiplying the ($k_x-k_y-\Omega$) Fourier amplitudes of the surface elevation by the following filter:
\begin{align}\label{filter}
f(\mathbf{k},\Omega)=f(k_x,k_y,\omega)=\begin{cases}
1, & |\Omega-\omega_{\mathbf{k}}|\leq \Delta\omega\\
0, & |\Omega-\omega_{\mathbf{k}}|> \Delta\omega,
\end{cases}
\end{align}
where $\Delta \omega$ is an appropriate filter width. In the presented analysis, we set the width of the filter $\Delta\omega=0.01$ rad/s. The filtered data $\hat{\eta}(x,y,t)$ are obtained starting from the surface elevation $\eta(x,y,t)$, computing its three dimensional Fourier transform, multiplying it by the filter $f(k_x,k_y,\omega)$, and finally computing its three dimensional inverse Fourier transform
\begin{align}\label{filteredeta}
    \hat{\eta}(x,y,t)=\mathscr{F}_{3D}^{-1}\left\{\mathscr{F}_{3D}\left[\eta(x,y,t)\right]f(k_x,k_y,\omega)\right\}.
\end{align}
This filtering procedure removes the bound modes from our data; such a procedure is analogous to the construction of the variable of interest $b_{\mathbf{k}}\left(t\right)$ through a near-identity transformation. 
Once $b_{\mathbf{k}}\left(t\right)$ is evaluated, the four-point correlator \eqref{c2} can be evaluated.\\

\subsection*{Analytical evaluation of the fourth-order correlator}

The evolution equation for $a_{\mathbf{k}}(t)$ in Eq.~\eqref{ak} contains all powers of nonlinearity, see \cite{krasitskii1994reduced,janssen2004interaction}. However, because of the absence of resonances in quadratic and some cubic terms, a near-identity canonical transformation (normal form expansion) can be performed to remove non-resonant triads and some quartets  \cite{janssen2009some}. The transformation is given by 
\begin{align}
a_{\mathbf{k}_1}=b_{\mathbf{k}_1}+\epsilon\sum_{\mathbf{k}_2,\mathbf{k}_3}\Big[& A_{\mathbf{k}_1,\mathbf{k}_2,\mathbf{k}_3}^{(1)}a_{\mathbf{k}_2}a_{\mathbf{k}_3}\delta_{\mathbf{k}_1,\mathbf{k}_2+\mathbf{k}_3}+A_{\mathbf{k}_1,\mathbf{k}_2,\mathbf{k}_3}^{(2)} a_{\mathbf{k}_2}^*a_{\mathbf{k}_3}\delta_{\mathbf{k}_1,-\mathbf{k}_2+\mathbf{k}_3}\nonumber\\
&\left.+A_{\mathbf{k}_1,\mathbf{k}_2,\mathbf{k}_3}^{(3)}a_{\mathbf{k}_2}^*a_{\mathbf{k}_3}^*\delta_{\mathbf{k}_1,-\mathbf{k}_2-\mathbf{k}_3}\right]+\mathcal{O}(\epsilon^2),
\end{align}
where $A_{\mathbf{k}_1,\mathbf{k}_2,\mathbf{k}_3}^{(i)}, i=1,2,3$ are coefficients to be determined in such a way that non-resonant terms are eliminated; $\delta_{\mathbf{k}_m,\mathbf{k}_n}$ is the Kronecker delta equal to 1 when $\mathbf{k}_m=\mathbf{k}_n$, and 0 otherwise. The final deterministic equation in terms of the new variable $b_{\mathbf{k}}(t)$, known as the Zakharov equation, takes the form
\begin{equation}
i \frac{\textrm{d} b_{\mathbf{k}_1}}{\textrm{d} t}=\omega_{\mathbf{k}_1} b_{\mathbf{k}_1}+ \epsilon^2\sum_{{\bf k}_2,{\bf k}_3,{\bf k}_4} T_{\mathbf{k}_1\mathbf{k}_2\mathbf{k}_3\mathbf{k}_4}b_{\mathbf{k}_2}^*b_{\mathbf{k}_3} b_{\mathbf{k}_4}\delta_{{\bf k}_1+{\bf k}_2,{\bf k}_3+{\bf k}_4},   
\end{equation}
where $\epsilon^2$ is a small parameter related to the strength of the nonlinearity and steepness and $T_{{\bf k}_1 {\bf k}_2{\bf k}_3{\bf k}_4}$ is a coupling coefficient whose form can be found in \cite{Janssen2007}; $\delta_{{\bf k}_1+{\bf k}_2,{\bf k}_3+{\bf k}_4} $ is the Kronecker delta. 
We now assume that $b_{\bf k}(t)$ are random variables with amplitudes and phases independent and identically distributed; moreover,  phases are uniformly distributed in the interval $[0,2\pi)$. This prescribed statistics defines the ensemble of random initial data, with averaging denoted by $\langle...\rangle$. The wave action spectrum $\langle|b_{\mathbf{k}}|^2\rangle$ evolves according to
\begin{equation}\label{ZakMod}
 \frac{\textrm{d} \langle|b_{\mathbf{k}_1}|^2\rangle}{\textrm{d}t}=2 \epsilon^2
 \mathfrak{Im} \left[\sum_{{\bf k}_2,{\bf k}_3,{\bf k}_4} T_{\mathbf{k}_1\mathbf{k}_2\mathbf{k}_3\mathbf{k}_4}\langle b_{\mathbf{k}_1}^*b_{\mathbf{k}_2}^*b_{\mathbf{k}_3} b_{\mathbf{k}_4}\rangle \delta_{{\bf k}_1+{\bf k}_2,{\bf k}_3+{\bf k}_4} 
 \right],
\end{equation}
where $\mathfrak{Im}$ stands for the imaginary part. {\color{black}One important aspect} of the Wave Turbulence Theory \cite{zakharov2012kolmogorov} is the estimation of the fourth-order correlator. A direct application of the Wick selection rule leads to the trivial result that the wave action does not evolve. Therefore, a higher-order closure is needed. The procedure leads to the following relation:
\begin{align}
    \langle &b_1^*(t)b_2^*(t)b_3(t)b_4(t)\rangle=\langle|\bar{b}_3|^2\rangle\langle|\bar{b}_4|^2\rangle\left(\delta_{3}^{1}\delta_{4}^{2}+\delta_{3}^{2}\delta_{4}^{1}\right)\nonumber \\
    &+2\epsilon^2 T_{1234}\langle|\bar{b}_1|^2\rangle\langle|\bar{b}_2|^2\rangle\langle|\bar{b}_3|^2\rangle\langle|\bar{b}_4|^2\rangle\left\{\frac{1}{\langle|\bar{b}_1|^2\rangle}+\frac{1}{\langle|\bar{b}_2|^2\rangle}-\frac{1}{\langle|\bar{b}_3|^2\rangle}-\frac{1}{\langle|\bar{b}_4|^2\rangle}\right\}\times \nonumber\\
&\times\frac{1-e^{-i\Delta\omega_{12}^{34}t}}{\Delta\omega_{12}^{34}}\delta_{12}^{34}.
\end{align}
After taking the large box and large time limits, and remembering that 
\begin{align}
\lim_{t\to+\infty}\frac{\sin\left(\Delta\omega_{12}^{34}t\right)}{\Delta\omega_{12}^{34}}=\pi\delta\left(\Delta\omega_{12}^{34}\right),
\end{align}
 we obtain for the imaginary part the desired equation (\ref{eq:6}).

\section*{Data availability}

All data supporting the findings of this study are available from the corresponding authors upon request.

\section*{Code availability}

The codes for the analysis of this study are available from the corresponding authors upon request.


\begin{thebibliography}{42}
\ifx \bisbn   \undefined \def \bisbn  #1{ISBN #1}\fi
\ifx \binits  \undefined \def \binits#1{#1}\fi
\ifx \bauthor  \undefined \def \bauthor#1{#1}\fi
\ifx \batitle  \undefined \def \batitle#1{#1}\fi
\ifx \bjtitle  \undefined \def \bjtitle#1{#1}\fi
\ifx \bvolume  \undefined \def \bvolume#1{\textbf{#1}}\fi
\ifx \byear  \undefined \def \byear#1{#1}\fi
\ifx \bissue  \undefined \def \bissue#1{#1}\fi
\ifx \bfpage  \undefined \def \bfpage#1{#1}\fi
\ifx \blpage  \undefined \def \blpage #1{#1}\fi
\ifx \burl  \undefined \def \burl#1{\textsf{#1}}\fi
\ifx \doiurl  \undefined \def \doiurl#1{\url{https://doi.org/#1}}\fi
\ifx \betal  \undefined \def \betal{\textit{et al.}}\fi
\ifx \binstitute  \undefined \def \binstitute#1{#1}\fi
\ifx \binstitutionaled  \undefined \def \binstitutionaled#1{#1}\fi
\ifx \bctitle  \undefined \def \bctitle#1{#1}\fi
\ifx \beditor  \undefined \def \beditor#1{#1}\fi
\ifx \bpublisher  \undefined \def \bpublisher#1{#1}\fi
\ifx \bbtitle  \undefined \def \bbtitle#1{#1}\fi
\ifx \bedition  \undefined \def \bedition#1{#1}\fi
\ifx \bseriesno  \undefined \def \bseriesno#1{#1}\fi
\ifx \blocation  \undefined \def \blocation#1{#1}\fi
\ifx \bsertitle  \undefined \def \bsertitle#1{#1}\fi
\ifx \bsnm \undefined \def \bsnm#1{#1}\fi
\ifx \bsuffix \undefined \def \bsuffix#1{#1}\fi
\ifx \bparticle \undefined \def \bparticle#1{#1}\fi
\ifx \barticle \undefined \def \barticle#1{#1}\fi
\bibcommenthead
\ifx \bconfdate \undefined \def \bconfdate #1{#1}\fi
\ifx \botherref \undefined \def \botherref #1{#1}\fi
\ifx \url \undefined \def \url#1{\textsf{#1}}\fi
\ifx \bchapter \undefined \def \bchapter#1{#1}\fi
\ifx \bbook \undefined \def \bbook#1{#1}\fi
\ifx \bcomment \undefined \def \bcomment#1{#1}\fi
\ifx \oauthor \undefined \def \oauthor#1{#1}\fi
\ifx \citeauthoryear \undefined \def \citeauthoryear#1{#1}\fi
\ifx \endbibitem  \undefined \def \endbibitem {}\fi
\ifx \bconflocation  \undefined \def \bconflocation#1{#1}\fi
\ifx \arxivurl  \undefined \def \arxivurl#1{\textsf{#1}}\fi
\csname PreBibitemsHook\endcsname

\bibitem[\protect\citeauthoryear{Komen et~al.}{1996}]{komen1996dynamics}
\begin{bbook}
\bauthor{\bsnm{Komen}, \binits{G.J.}},
\bauthor{\bsnm{Cavaleri}, \binits{L.}},
\bauthor{\bsnm{Donelan}, \binits{M.}},
\bauthor{\bsnm{Hasselmann}, \binits{K.}},
\bauthor{\bsnm{Hasselmann}, \binits{S.}},
\bauthor{\bsnm{Janssen}, \binits{P.}}:
\bbtitle{Dynamics and Modelling of Ocean Waves},
(\byear{1996})
\end{bbook}
\endbibitem

\bibitem[\protect\citeauthoryear{Janssen}{2004}]{janssen2004interaction}
\begin{bbook}
\bauthor{\bsnm{Janssen}, \binits{P.}}:
\bbtitle{The Interaction of Ocean Waves and Wind}.
\bpublisher{Cambridge University Press},
\blocation{Cambridge}
(\byear{2004})
\end{bbook}
\endbibitem

\bibitem[\protect\citeauthoryear{Phillips}{1960}]{phillips1960dynamics}
\begin{barticle}
\bauthor{\bsnm{Phillips}, \binits{O.}}:
\batitle{On the dynamics of unsteady gravity waves of finite amplitude part 1.
  the elementary interactions}.
\bjtitle{Journal of Fluid Mechanics}
\bvolume{9}(\bissue{2}),
\bfpage{193}--\blpage{217}
(\byear{1960})
\end{barticle}
\endbibitem

\bibitem[\protect\citeauthoryear{Janssen}{2009}]{janssen2009some}
\begin{barticle}
\bauthor{\bsnm{Janssen}, \binits{P.A.}}:
\batitle{On some consequences of the canonical transformation in the
  hamiltonian theory of water waves}.
\bjtitle{Journal of Fluid Mechanics}
\bvolume{637},
\bfpage{1}--\blpage{44}
(\byear{2009})
\end{barticle}
\endbibitem

\bibitem[\protect\citeauthoryear{Zakharov
  et~al.}{2012}]{zakharov2012kolmogorov}
\begin{bbook}
\bauthor{\bsnm{Zakharov}, \binits{V.E.}},
\bauthor{\bsnm{L'vov}, \binits{V.S.}},
\bauthor{\bsnm{Falkovich}, \binits{G.}}:
\bbtitle{Kolmogorov Spectra of Turbulence I: Wave Turbulence}.
\bpublisher{Springer}, \blocation{Cambridge}
(\byear{2012})
\end{bbook}
\endbibitem

\bibitem[\protect\citeauthoryear{Krasitskii}{1994}]{krasitskii1994reduced}
\begin{barticle}
\bauthor{\bsnm{Krasitskii}, \binits{V.P.}}:
\batitle{On reduced equations in the hamiltonian theory of weakly nonlinear
  surface waves}.
\bjtitle{Journal of Fluid Mechanics}
\bvolume{272},
\bfpage{1}--\blpage{20}
(\byear{1994})
\end{barticle}
\endbibitem

\bibitem[\protect\citeauthoryear{Longuet-Higgins and
  Smith}{1966}]{longuet1966experiment}
\begin{barticle}
\bauthor{\bsnm{Longuet-Higgins}, \binits{M.}},
\bauthor{\bsnm{Smith}, \binits{N.}}:
\batitle{An experiment on third-order resonant wave interactions}.
\bjtitle{Journal of Fluid Mechanics}
\bvolume{25}(\bissue{3}),
\bfpage{417}--\blpage{435}
(\byear{1966})
\end{barticle}
\endbibitem

\bibitem[\protect\citeauthoryear{McGoldrick
  et~al.}{1966}]{mcgoldrick1966measurements}
\begin{barticle}
\bauthor{\bsnm{McGoldrick}, \binits{L.}},
\bauthor{\bsnm{Phillips}, \binits{O.}},
\bauthor{\bsnm{Huang}, \binits{N.}},
\bauthor{\bsnm{Hodgson}, \binits{T.}}:
\batitle{Measurements of third-order resonant wave interactions}.
\bjtitle{Journal of Fluid Mechanics}
\bvolume{25}(\bissue{3}),
\bfpage{437}--\blpage{456}
(\byear{1966})
\end{barticle}
\endbibitem

\bibitem[\protect\citeauthoryear{Waseda et~al.}{2015}]{waseda2015third}
\begin{barticle}
\bauthor{\bsnm{Waseda}, \binits{T.}},
\bauthor{\bsnm{Kinoshita}, \binits{T.}},
\bauthor{\bsnm{Cavaleri}, \binits{L.}},
\bauthor{\bsnm{Toffoli}, \binits{A.}}:
\batitle{Third-order resonant wave interactions under the influence of
  background current fields}.
\bjtitle{Journal of Fluid Mechanics}
\bvolume{784},
\bfpage{51}--\blpage{73}
(\byear{2015})
\end{barticle}
\endbibitem

\bibitem[\protect\citeauthoryear{Bonnefoy
  et~al.}{2016}]{bonnefoy2016observation}
\begin{barticle}
\bauthor{\bsnm{Bonnefoy}, \binits{F.}},
\bauthor{\bsnm{Haudin}, \binits{F.}},
\bauthor{\bsnm{Michel}, \binits{G.}},
\bauthor{\bsnm{Semin}, \binits{B.}},
\bauthor{\bsnm{Humbert}, \binits{T.}},
\bauthor{\bsnm{Auma{\^\i}tre}, \binits{S.}},
\bauthor{\bsnm{Berhanu}, \binits{M.}},
\bauthor{\bsnm{Falcon}, \binits{E.}}:
\batitle{Observation of resonant interactions among surface gravity waves}.
\bjtitle{Journal of Fluid Mechanics}
\bvolume{805},
\bfpage{3}
(\byear{2016})
\end{barticle}
\endbibitem

\bibitem[\protect\citeauthoryear{Hasselmann}{1962}]{hasselmann1962non}
\begin{barticle}
\bauthor{\bsnm{Hasselmann}, \binits{K.}}:
\batitle{On the non-linear energy transfer in a gravity-wave spectrum part 1.
  general theory}.
\bjtitle{Journal of Fluid Mechanics}
\bvolume{12}(\bissue{4}),
\bfpage{481}--\blpage{500}
(\byear{1962})
\end{barticle}
\endbibitem

\bibitem[\protect\citeauthoryear{Zakharov and
  Filonenko}{1966}]{zakharov1966energy}
\begin{bchapter}
\bauthor{\bsnm{Zakharov}, \binits{V.E.}},
\bauthor{\bsnm{Filonenko}, \binits{N.}}:
\bctitle{Energy spectrum for stochastic oscillations of the surface of a
  liquid}.
In: \bbtitle{Doklady Akademii Nauk},
vol. \bseriesno{170},
pp. \bfpage{1292}--\blpage{1295}
(\byear{1966}).
\bcomment{Russian Academy of Sciences}
\end{bchapter}
\endbibitem

\bibitem[\protect\citeauthoryear{Cavaleri et~al.}{2007}]{cavaleri2007wave}
\begin{barticle}
\bauthor{\bsnm{Cavaleri}, \binits{L.}},
\bauthor{\bsnm{Alves}, \binits{J.-H.}},
\bauthor{\bsnm{Ardhuin}, \binits{F.}},
\bauthor{\bsnm{Babanin}, \binits{A.}},
\bauthor{\bsnm{Banner}, \binits{M.}},
\bauthor{\bsnm{Belibassakis}, \binits{K.}},
\bauthor{\bsnm{Benoit}, \binits{M.}},
\bauthor{\bsnm{Donelan}, \binits{M.}},
\bauthor{\bsnm{Groeneweg}, \binits{J.}},
\bauthor{\bsnm{Herbers}, \binits{T.}}, \betal:
\batitle{Wave modelling--the state of the art}.
\bjtitle{Progress in oceanography}
\bvolume{75}(\bissue{4}),
\bfpage{603}--\blpage{674}
(\byear{2007})
\end{barticle}
\endbibitem

\bibitem[\protect\citeauthoryear{Benetazzo}{2006}]{benetazzo2006measurements}
\begin{barticle}
\bauthor{\bsnm{Benetazzo}, \binits{A.}}:
\batitle{Measurements of short water waves using stereo matched image
  sequences}.
\bjtitle{Coastal engineering}
\bvolume{53}(\bissue{12}),
\bfpage{1013}--\blpage{1032}
(\byear{2006})
\end{barticle}
\endbibitem

\bibitem[\protect\citeauthoryear{Benetazzo
  et~al.}{2012}]{benetazzo2012offshore}
\begin{barticle}
\bauthor{\bsnm{Benetazzo}, \binits{A.}},
\bauthor{\bsnm{Fedele}, \binits{F.}},
\bauthor{\bsnm{Gallego}, \binits{G.}},
\bauthor{\bsnm{Shih}, \binits{P.-C.}},
\bauthor{\bsnm{Yezzi}, \binits{A.}}:
\batitle{Offshore stereo measurements of gravity waves}.
\bjtitle{Coastal Engineering}
\bvolume{64},
\bfpage{127}--\blpage{138}
(\byear{2012})
\end{barticle}
\endbibitem

\bibitem[\protect\citeauthoryear{Romero and
  Melville}{2010}]{AirborneObservationsofFetchLimitedWavesintheGulfofTehuantepec}
\begin{barticle}
\bauthor{\bsnm{Romero}, \binits{L.}},
\bauthor{\bsnm{Melville}, \binits{W.K.}}:
\batitle{Airborne observations of fetch-limited waves in the gulf of
  tehuantepec}.
\bjtitle{Journal of Physical Oceanography}
\bvolume{40}(\bissue{3}),
\bfpage{441}--\blpage{465}
(\byear{2010})
\doiurl{10.1175/2009JPO4127.1}
\end{barticle}
\endbibitem

\bibitem[\protect\citeauthoryear{Lenain and
  Melville}{2017}]{lenain2017measurements}
\begin{barticle}
\bauthor{\bsnm{Lenain}, \binits{L.}},
\bauthor{\bsnm{Melville}, \binits{W.K.}}:
\batitle{Measurements of the directional spectrum across the equilibrium
  saturation ranges of wind-generated surface waves}.
\bjtitle{Journal of Physical Oceanography}
\bvolume{47}(\bissue{8}),
\bfpage{2123}--\blpage{2138}
(\byear{2017})
\end{barticle}
\endbibitem

\bibitem[\protect\citeauthoryear{Kawai et~al.}{1977}]{kawai1977field}
\begin{barticle}
\bauthor{\bsnm{Kawai}, \binits{S.}},
\bauthor{\bsnm{Okada}, \binits{K.}},
\bauthor{\bsnm{Toba}, \binits{Y.}}:
\batitle{Field data support of three-seconds power law and gu* $\sigma$-
  4-spectral form for growing wind waves}.
\bjtitle{Journal of the Oceanographical Society of Japan}
\bvolume{33},
\bfpage{137}--\blpage{150}
(\byear{1977})
\end{barticle}
\endbibitem

\bibitem[\protect\citeauthoryear{Hwang et~al.}{2000}]{hwang2000airborne}
\begin{barticle}
\bauthor{\bsnm{Hwang}, \binits{P.A.}},
\bauthor{\bsnm{Wang}, \binits{D.W.}},
\bauthor{\bsnm{Walsh}, \binits{E.J.}},
\bauthor{\bsnm{Krabill}, \binits{W.B.}},
\bauthor{\bsnm{Swift}, \binits{R.N.}}:
\batitle{Airborne measurements of the wavenumber spectra of ocean surface
  waves. part i: Spectral slope and dimensionless spectral coefficient}.
\bjtitle{Journal of physical oceanography}
\bvolume{30}(\bissue{11}),
\bfpage{2753}--\blpage{2767}
(\byear{2000})
\end{barticle}
\endbibitem

\bibitem[\protect\citeauthoryear{Resio et~al.}{2004}]{resio2004equilibrium}
\begin{botherref}
\oauthor{\bsnm{Resio}, \binits{D.T.}},
\oauthor{\bsnm{Long}, \binits{C.E.}},
\oauthor{\bsnm{Vincent}, \binits{C.L.}}:
Equilibrium-range constant in wind-generated wave spectra.
Journal of Geophysical Research: Oceans
\textbf{109}(C1)
(2004)
\end{botherref}
\endbibitem

\bibitem[\protect\citeauthoryear{Ewans}{1998}]{ewans1998observations}
\begin{barticle}
\bauthor{\bsnm{Ewans}, \binits{K.C.}}:
\batitle{Observations of the directional spectrum of fetch-limited waves}.
\bjtitle{Journal of Physical Oceanography}
\bvolume{28}(\bissue{3}),
\bfpage{495}--\blpage{512}
(\byear{1998})
\end{barticle}
\endbibitem

\bibitem[\protect\citeauthoryear{Romero and
  Melville}{2010}]{romero2010airborne}
\begin{barticle}
\bauthor{\bsnm{Romero}, \binits{L.}},
\bauthor{\bsnm{Melville}, \binits{W.K.}}:
\batitle{Airborne observations of fetch-limited waves in the gulf of
  tehuantepec}.
\bjtitle{Journal of Physical Oceanography}
\bvolume{40}(\bissue{3}),
\bfpage{441}--\blpage{465}
(\byear{2010})
\end{barticle}
\endbibitem

\bibitem[\protect\citeauthoryear{Badulin and Zakharov}{2017}]{badulin2017ocean}
\begin{barticle}
\bauthor{\bsnm{Badulin}, \binits{S.I.}},
\bauthor{\bsnm{Zakharov}, \binits{V.E.}}:
\batitle{Ocean swell within the kinetic equation for water waves}.
\bjtitle{Nonlinear Processes in Geophysics}
\bvolume{24}(\bissue{2}),
\bfpage{237}--\blpage{253}
(\byear{2017})
\end{barticle}
\endbibitem

\bibitem[\protect\citeauthoryear{Toffoli et~al.}{2010}]{toffoli2010development}
\begin{botherref}
\oauthor{\bsnm{Toffoli}, \binits{A.}},
\oauthor{\bsnm{Onorato}, \binits{M.}},
\oauthor{\bsnm{Bitner-Gregersen}, \binits{E.}},
\oauthor{\bsnm{Monbaliu}, \binits{J.}}:
Development of a bimodal structure in ocean wave spectra.
Journal of Geophysical Research: Oceans
\textbf{115}(C3)
(2010)
\end{botherref}
\endbibitem

\bibitem[\protect\citeauthoryear{Phillips}{1985}]{phillips1985spectral}
\begin{barticle}
\bauthor{\bsnm{Phillips}, \binits{O.}}:
\batitle{Spectral and statistical properties of the equilibrium range in
  wind-generated gravity waves}.
\bjtitle{Journal of Fluid Mechanics}
\bvolume{156},
\bfpage{505}--\blpage{531}
(\byear{1985})
\end{barticle}
\endbibitem

\bibitem[\protect\citeauthoryear{Romero et~al.}{2012}]{romero2012spectral}
\begin{barticle}
\bauthor{\bsnm{Romero}, \binits{L.}},
\bauthor{\bsnm{Melville}, \binits{W.K.}},
\bauthor{\bsnm{Kleiss}, \binits{J.M.}}:
\batitle{Spectral energy dissipation due to surface wave breaking}.
\bjtitle{Journal of Physical Oceanography}
\bvolume{42}(\bissue{9}),
\bfpage{1421}--\blpage{1444}
(\byear{2012})
\end{barticle}
\endbibitem

\bibitem[\protect\citeauthoryear{Phillips}{1958}]{phillips1958equilibrium}
\begin{barticle}
\bauthor{\bsnm{Phillips}, \binits{O.M.}}:
\batitle{The equilibrium range in the spectrum of wind-generated waves}.
\bjtitle{Journal of Fluid Mechanics}
\bvolume{4}(\bissue{4}),
\bfpage{426}--\blpage{434}
(\byear{1958})
\end{barticle}
\endbibitem

\bibitem[\protect\citeauthoryear{Romero}{2019}]{romero2019distribution}
\begin{barticle}
\bauthor{\bsnm{Romero}, \binits{L.}}:
\batitle{Distribution of surface wave breaking fronts}.
\bjtitle{Geophysical Research Letters}
\bvolume{46}(\bissue{17-18}),
\bfpage{10463}--\blpage{10474}
(\byear{2019})
\end{barticle}
\endbibitem

\bibitem[\protect\citeauthoryear{Akaawase
  et~al.}{2025}]{akaawase2025observations}
\begin{barticle}
\bauthor{\bsnm{Akaawase}, \binits{B.}},
\bauthor{\bsnm{Romero}, \binits{L.}},
\bauthor{\bsnm{Benetazzo}, \binits{A.}}:
\batitle{Observations of wave-breaking direction and energy spread}.
\bjtitle{Geophysical Research Letters}
\bvolume{52}(\bissue{14}),
\bfpage{2025}--\blpage{116452}
(\byear{2025})
\end{barticle}
\endbibitem

\bibitem[\protect\citeauthoryear{Campagne
  et~al.}{2019}]{campagne2019identifying}
\begin{barticle}
\bauthor{\bsnm{Campagne}, \binits{A.}},
\bauthor{\bsnm{Hassaini}, \binits{R.}},
\bauthor{\bsnm{Redor}, \binits{I.}},
\bauthor{\bsnm{Valran}, \binits{T.}},
\bauthor{\bsnm{Viboud}, \binits{S.}},
\bauthor{\bsnm{Sommeria}, \binits{J.}},
\bauthor{\bsnm{Mordant}, \binits{N.}}:
\batitle{Identifying four-wave-resonant interactions in a surface gravity wave
  turbulence experiment}.
\bjtitle{Physical Review Fluids}
\bvolume{4}(\bissue{7}),
\bfpage{074801}
(\byear{2019})
\end{barticle}
\endbibitem

\bibitem[\protect\citeauthoryear{Zakharov and Filonenko}{1967}]{Zakharov1967}
\begin{barticle}
\bauthor{\bsnm{Zakharov}, \binits{V.}},
\bauthor{\bsnm{Filonenko}, \binits{N.}}:
\batitle{Weak turbulence of capillary waves}.
\bjtitle{Journal of Applied Mechanics and Technical Physics}
\bvolume{8}(\bissue{5}),
\bfpage{37}--\blpage{40}
(\byear{1967})
\doiurl{10.1007/BF00915178}
\end{barticle}
\endbibitem

\bibitem[\protect\citeauthoryear{Benney}{1962}]{benney1962non}
\begin{barticle}
\bauthor{\bsnm{Benney}, \binits{D.}}:
\batitle{Non-linear gravity wave interactions}.
\bjtitle{Journal of Fluid Mechanics}
\bvolume{14}(\bissue{4}),
\bfpage{577}--\blpage{584}
(\byear{1962})
\end{barticle}
\endbibitem

\bibitem[\protect\citeauthoryear{Benney and
  Saffman}{1966}]{benney1966nonlinear}
\begin{barticle}
\bauthor{\bsnm{Benney}, \binits{D.}},
\bauthor{\bsnm{Saffman}, \binits{P.G.}}:
\batitle{Nonlinear interactions of random waves in a dispersive medium}.
\bjtitle{Proceedings of the Royal Society of London. Series A. Mathematical and
  Physical Sciences}
\bvolume{289}(\bissue{1418}),
\bfpage{301}--\blpage{320}
(\byear{1966})
\end{barticle}
\endbibitem

\bibitem[\protect\citeauthoryear{Benney and Newell}{1969}]{benney1969random}
\begin{barticle}
\bauthor{\bsnm{Benney}, \binits{D.}},
\bauthor{\bsnm{Newell}, \binits{A.C.}}:
\batitle{Random wave closures}.
\bjtitle{Studies in Applied Mathematics}
\bvolume{48}(\bissue{1}),
\bfpage{29}--\blpage{53}
(\byear{1969})
\end{barticle}
\endbibitem

\bibitem[\protect\citeauthoryear{Longuet-Higgins}{1962}]{longuet1962resonant}
\begin{barticle}
\bauthor{\bsnm{Longuet-Higgins}, \binits{M.S.}}:
\batitle{Resonant interactions between two trains of gravity waves}.
\bjtitle{Journal of Fluid Mechanics}
\bvolume{12}(\bissue{3}),
\bfpage{321}--\blpage{332}
(\byear{1962})
\end{barticle}
\endbibitem

\bibitem[\protect\citeauthoryear{Annenkov and Shrira}{2018}]{annenkov2018}
\begin{barticle}
\bauthor{\bsnm{Annenkov}, \binits{S.Y.}},
\bauthor{\bsnm{Shrira}, \binits{V.I.}}:
\batitle{Spectral evolution of weakly nonlinear random waves: kinetic
  description versus direct numerical simulations}.
\bjtitle{Journal of Fluid Mechanics}
\bvolume{844},
\bfpage{766}--\blpage{795}
(\byear{2018})
\end{barticle}
\endbibitem

\bibitem[\protect\citeauthoryear{Guimar{\~a}es et~al.}{2020}]{Guimaraes2020}
\begin{barticle}
\bauthor{\bsnm{Guimar{\~a}es}, \binits{P.V.}},
\bauthor{\bsnm{Ardhuin}, \binits{F.}},
\bauthor{\bsnm{Bergamasco}, \binits{F.}},
\bauthor{\bsnm{Leckler}, \binits{F.}},
\bauthor{\bsnm{Filipot}, \binits{J.-F.}},
\bauthor{\bsnm{Shim}, \binits{J.-S.}},
\bauthor{\bsnm{Dulov}, \binits{V.}},
\bauthor{\bsnm{Benetazzo}, \binits{A.}}:
\batitle{A data set of sea surface stereo images to resolve space-time wave
  fields}.
\bjtitle{Scientific Data}
\bvolume{7}(\bissue{1}),
\bfpage{145}
(\byear{2020})
\doiurl{10.1038/s41597-020-0492-9}
\end{barticle}
\endbibitem

\bibitem[\protect\citeauthoryear{Bergamasco et~al.}{2017}]{bergamasco2017wass}
\begin{barticle}
\bauthor{\bsnm{Bergamasco}, \binits{F.}},
\bauthor{\bsnm{Torsello}, \binits{A.}},
\bauthor{\bsnm{Sclavo}, \binits{M.}},
\bauthor{\bsnm{Barbariol}, \binits{F.}},
\bauthor{\bsnm{Benetazzo}, \binits{A.}}:
\batitle{Wass: An open-source pipeline for 3d stereo reconstruction of ocean
  waves}.
\bjtitle{Computers \& Geosciences}
\bvolume{107},
\bfpage{28}--\blpage{36}
(\byear{2017})
\end{barticle}
\endbibitem

\bibitem[\protect\citeauthoryear{Zhang and Pan}{2022}]{Zhang2022}
\begin{barticle}
\bauthor{\bsnm{Zhang}, \binits{Z.}},
\bauthor{\bsnm{Pan}, \binits{Y.}}:
\batitle{Numerical investigation of turbulence of surface gravity waves}.
\bjtitle{Journal of Fluid Mechanics}
\bvolume{933},
\bfpage{58}
(\byear{2022})
\doiurl{10.1017/jfm.2021.1114}
\end{barticle}
\endbibitem

\bibitem[\protect\citeauthoryear{Onorato et~al.}{2013}]{onorato2013rogue}
\begin{barticle}
\bauthor{\bsnm{Onorato}, \binits{M.}},
\bauthor{\bsnm{Residori}, \binits{S.}},
\bauthor{\bsnm{Bortolozzo}, \binits{U.}},
\bauthor{\bsnm{Montina}, \binits{A.}},
\bauthor{\bsnm{Arecchi}, \binits{F.}}:
\batitle{Rogue waves and their generating mechanisms in different physical
  contexts}.
\bjtitle{Physics Reports}
\bvolume{528}(\bissue{2}),
\bfpage{47}--\blpage{89}
(\byear{2013})
\end{barticle}
\endbibitem

\bibitem[\protect\citeauthoryear{Young}{1999}]{young1999wind}
\begin{bbook}
\bauthor{\bsnm{Young}, \binits{I.R.}}:
\bbtitle{Wind Generated Ocean Waves}
vol. \bseriesno{2}.
\bpublisher{Elsevier}, \blocation{Oxford}
(\byear{1999})
\end{bbook}
\endbibitem

\bibitem[\protect\citeauthoryear{Janssen and Onorato}{2007}]{Janssen2007}
\begin{barticle}
\bauthor{\bsnm{Janssen}, \binits{P.}},
\bauthor{\bsnm{Onorato}, \binits{M.}}:
\batitle{The intermediate water depth limit of the zakharov equation and
  consequences for wave prediction}.
\bjtitle{Journal of Physical Oceanography}
\bvolume{37},
\bfpage{2389}--\blpage{2400}
(\byear{2007})
\doiurl{10.1175/JPO3128.1}
\end{barticle}
\endbibitem

\end{thebibliography}


\section*{Acknowledgments}
The authors would like to thank P. Lizzio and A. Muratori for their help during the early stages of the work. This research was supported by Simons Collaboration on Wave Turbulence, Grant
No. 617006. M.O. is also supported by INFN (MMNLP and FieldTurb)

\section*{Author contributions}
DM, GD, AB, and MO designed the research, performed the analysis, and wrote the manuscript.
Correspondence to giovannidematteis@gmail.com

\section*{Competing interests}
The authors declare no competing interests.

\end{document}